\documentclass[aip,rsi,reprint, floatfix]{revtex4-1} 

\usepackage{amsmath}
\usepackage{graphicx}
\usepackage{xfrac}

\graphicspath{{Figs/}}

\draft 

\begin{document}


\title{A low noise modular current source for stable magnetic field control} 



\author{Valerio Biancalana}
\author{Giuseppe Bevilacqua}
\author{Piero Chessa}
\affiliation{DIISM University of Siena; CNISM. Via Roma, 56 -- 53100 Siena, Italy}


\author{Yordanka Dancheva}

\author{Roberto Cecchi}
\author{Leonardo Stiaccini}
\affiliation{DSFTA University of Siena. Via Roma, 56 -- 53100 Siena, Italy}


\date{\today}

\begin{abstract}
\footnotesize
A low cost, stable, programmable, unipolar current source is described. The circuit is designed in view of a modular arrangement, suitable for applications where several DC sources must be controlled at once. A hybrid switching/linear design helps in improving the stability and in reducing the power dissipation and cross-talking. Multiple units can be supplied by a single DC power supply, while allowing for a variety of maximal current values and compliance voltages at the outputs. The circuit is analogically controlled by an unipolar voltage, enabling current programmability and control through commercial digital-to-analogue conversion cards.
\end{abstract}

 


\pacs{07.50.-e, 07.50.Ek, 07.55.Ge}

\maketitle 

\section{Introduction}
\label{sec:introduction}

In many laboratory applications precise electric current supplies are required \cite{skyba_rsi_91,  burin_rsi_96, linzen_rsi_04,  belfi_rsi_10, talukdar_rsi_11, scandurra_rsi_14}. Depending on the particular experiment, different kinds of specifications are necessary in terms of noise, long-term stability, power dissipation, slew rate, single or dual polarity, programmability, and so on. Excellent power supplies are commercially available, which allow for both voltage and current programming, within wide sets of ranges. Such a multi-purpose instrumentation may correspondingly be suitable to large varieties of requirements.
However, applications exist  where a relevant number of current sources are supposed to work within narrow and well defined ranges. This makes the commercial solution unnecessarily bulky and expensive.

As an example, in our laboratory we run an experiment \cite{bevilacqua_pra_16, bevilacqua_apb_16, bevilacqua_jmr_16} where eight low noise current sources are needed. They operate at different working points (currents have typical values ranging from  10 to 1000~mA), are applied to different loads (typical load voltages range from 5 to 25~V), and are to be adjusted within narrow ranges (the relative current adjustments rarely exceed a few percent). There are evident advantages in controlling their outputs via a custom computer program. The numeric control also enables the optimal current outputs to be determined automatically.

In this note a modular current source is proposed and described. A set made of eight units is built, with coarse hardware setting of the operating range and fine software setting of the working point. The whole set requires a single fixed DC voltage power supply, but the compliance voltages at the linear stage outputs are dynamically adjusted on the basis of the load specifications. The power supply output must exceed by a few volt (3V at least) the highest voltage drop in the loads. Such a kind of solution represents an optimum in terms of costs, reliability, power consumption, simplicity in implementation and setting-up. The drop voltage over the linear stage is kept at a roughly constant level, which helps in reducing thermal drifts and residual cross-talking.

\section {Circuit}
\label{sec:circuit}

Hybrid power supplies constitute an appealing and obvious solution when searching for a compromise between low power dissipation and low ripple \cite{keeping_digikey_12}. The residual ripple produced by switching regulators can be profitably reduced by a final linear regulation stage. The switching regulator output contains a ripple component, often coupled with periodic spikes. In the frequency domain, the ripple appears as a peak at the switching frequency and by a number of secondary peaks at its harmonics, while the occasional spikes are responsible of high-order harmonics, which typically broaden the noise spectrum up to tens/hundreds of MHz.

\begin{figure} [htbp]\centering
 \includegraphics [width=0.9\columnwidth] {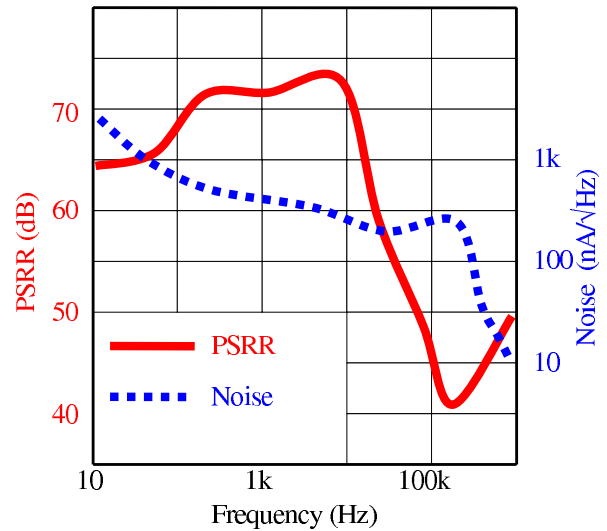}
 \caption{Typical behavior of PSRR (left scale, red), and current noise (right scale, blue) of a linear stage, as can be found in the literature and data-sheets. 
 \label{fig:psrrnoise}}%
\end{figure}

A linear stage helps in rejecting the ripple components, according to its so called PSRR (power supply ripple rejection). The PSRR of linear stages generally behaves  as qualitatively represented in Fig.\ref{fig:psrrnoise}.

As shown, linear stages
 feature an optimal PSRR at intermediate frequencies, while the suppression of the high frequency terms caused by the spikes and higher harmonics of the ripple is less effective due to the high frequency cut-off in the circuitry response, which is in turn set by the gain-bandwidth product of the error amplifier. However, in those cases where the current source supplies inductive loads, the reactance acts to filter out  the ripple components and the spike/high frequency terms.

At the other side of the spectrum, i.e. at very low frequencies, the linear stage typically displays a worse PSRR, as well. Thus, the low frequency noise components of the power supply are barely attenuated on the load. In addition, the linear stage introduces an intrinsic disturbance that increases at low frequency  (1/f noise), as qualitatively shown in the second plot (right scale) of Fig.\ref{fig:psrrnoise}. At very low frequencies, the noise introduced by the linear stage often becomes dominant. 

Information is available in the Application Notes released by  several producers, both accounting for the PSRR  and for the intrinsic noise introduced by the linear stage \cite{teel_ti_05}, particularly in the case when they operate in a hybrid setup \cite{keeping_digikey_15}.

Most noise patterns reported in literature are represented by spectral plots extending down to about 10~Hz - 1~Hz, while slower fluctuations are usually described in terms of drifts in the time domain.

Such drifts are caused by slowly varying, temperature dependent parameters\cite{linzen_rsi_04} and commonly appear at warming-up, after output current adjustments, and in consequence of environmental temperature variations and air flow. The latter are particularly important when having high power dissipation.

By coupling a linear with a switching stage working at intermediate frequencies, with passive low-pass filters stopping the higher harmonics, the high-frequency drop of the PSRR can be suitably counteracted. The switching stage reduces the power dissipation level facilitating the task of  keeping the circuit temperature constant, this enhances the long-term stability improving the spectral noise feature at very low frequencies.

A design based on hybrid implementation is important for modular systems. A single DC source supplies a set of current generators where all the linear stages work in a low drop-out condition, while the diverse output voltage requirements (depending on both current set-points and load impedance) are regulated at the preliminary switching stage.

Another important advantage introduced by the hybrid topology in modular system is the reduction of cross-talk effects. When the current driven by one unit is changed, some variation will occur in the output voltage of the main DC supply. The switching stages prevent such voltage step to appear as a drop-out voltage variation across the other linear stages, which would in turn result in an output current step, due to the finite zero-frequency PSRR.

We have designed, built and tested a hybrid circuit based on a commercial switching device and on a low-noise op-amp controlled linear stage capable of controlling output current within presettable ranges, up to 2~A. The full scale is set by selecting the value of a reference resistance. The drop-out voltage over the final transistor is dynamically stabilised, acting on the switching stage. Passive filters complete the scheme to filter out spiky terms. Care was taken in the circuit design to maximise the distance between heating devices (final transistor and switching chip) and the elements devoted to error signal detection and amplification.

A simple evaluation of the relative current stability $\delta I /I$ of a voltage controlled current source, taking into consideration the fluctuations $\delta R$ (mainly due to temperature drifts) of the reference resistance R and the ones $\delta V$ of the controlling signal $V$ and the of the error signal amplifiers, gives 
\begin {equation}
\frac{\delta I}{I}=\frac{\delta R}{R}+\frac{\delta V}{V}=\frac{\delta R}{R}+\frac{\delta V}{RI},
\end {equation}
from which a few hints are immediately inferred: R must be selected with very low temperature coefficient and placed at large distance from heating devices; bigger values of R are preferable, in order to operate at rather high controlling voltages.

\begin{figure} [htbp]\centering
 \includegraphics [width=0.9\columnwidth] {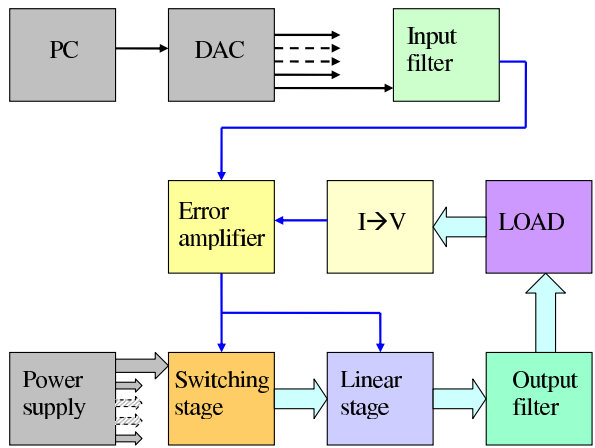}
 \caption{Block scheme of the modular set (in colour the part of a single unit). The dropout voltage on each linear stage is roughly stabilised with a feed-forward control of the switching stage. The linear stage elements (particularly the error amplifier and the $I \rightarrow V$ converter) are selected for a small offset and offset drift. A more detailed schematic is available as a supplemental material.
 \label{fig:blockscheme}}%
\end{figure}

Fig.\ref{fig:blockscheme} shows the block-scheme of a single unit. 
The reference voltage generated by the digital-to-analogue converter (DAC) is filtered, attenuated and sent to the non-inverting input of a low-offset operational amplifier belonging to the feedback loop.  The inverting input is set at a voltage proportional to the actual load current. The error signal, consisting of the difference between actual and set load current, is amplified and sent to a power amplifier with very low output impedance, made by two BJT, NPN transistors in Darlington configuration, whose output is applied to the load. 
The amplified error signal is also sent to the switching stage in a feed-forward chain, in such a way to adapt its output voltage to the linear stage. The collector-emitter junction of the final BJT is maintained at a low and roughly constant level (about 3V). Special care is devoted to filter the supply voltage of all the active elements in the feedback loop, with emphasis on the error amplifier and the power amplifier. It is possible to set the full range by changing the reference resistor that converts the load current into voltage. This enables a compromise between the maximal controllable current and the full exploitation of the DAC resolution to be found.

\section {Electrical characterization of the current supply}
\label{sec:characterization}

\begin{figure} [htbp]\centering
\includegraphics [width=0.9\columnwidth] {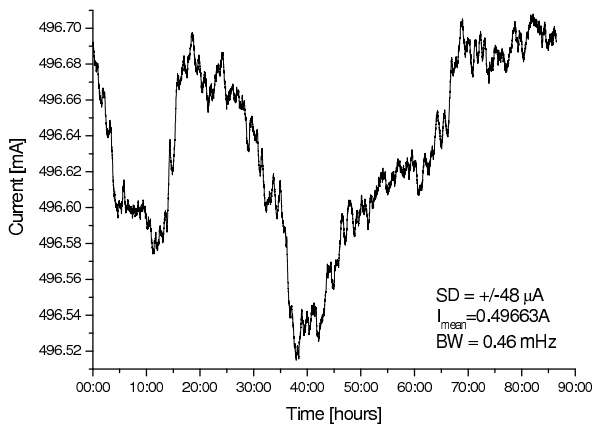}
\caption{Output current long term stability over hours. The current is sampled with 0.46~mHz rate. The load is a pair of large area, Helmholtz coils with 35~mH inductance  and 20~Ohm resistance. 
\label{fig:drift}}%
\end{figure}

To characterise one unit, the output current is measured by using a 6\sfrac{1}{2} digit multimeter. The output current noise spectra and the current drift over time intervals as long as tens of hours are measured. The cross-talk when using multiple units is evaluated as well.
Fig.\ref{fig:drift} shows the current stability at bias current of roughly 0.5~A on a 20~Ohm load (with a compliance voltage of 10~V). 
The observed long-term drift  is mainly caused by a daily temperature variation of several degrees Celsius. The drift pattern resulting from this measurements sets a relative uncertainty of $10^{-4}$, which is comparable with the relative uncertainty of the reference voltage (relative uncertainty of the DAC card used in this implementation). Thus, the unit output current stability is better than (or comparable with) the one given by the DAC.

\begin{figure} [htbp]\centering
\includegraphics [width=0.9\columnwidth] {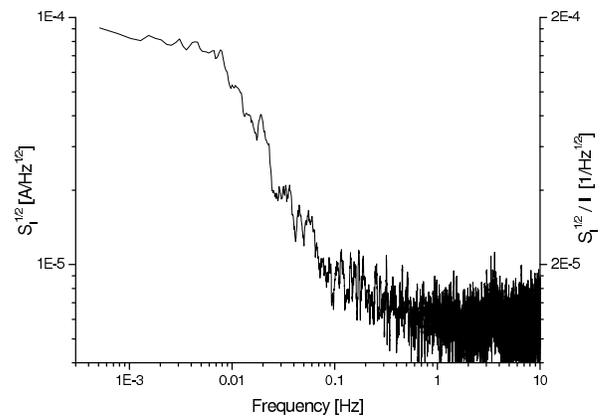}
\caption{Spectral density of the current noise.
\label{fig:noisepsd}}%
\end{figure}

Fig.\ref{fig:noisepsd} shows the power density of the current noise. The noise below 0.1~Hz has a $1/f^2$ power spectrum, which is characteristic for a Brownian noise, suggesting that the drift dynamics is driven by a Wiener process. No peaks at the switching frequency emerge from the noise floor in the spectral range not shown in that figure.

The cross talk in a set of eight units is estimated as follows: one unit is set at a 500~mA and the remaining units' outputs are set either to zero or to full range. The total current supplied by those seven units changes by 2.3~A and causes a 720~mV step in the output voltage of the main supply. Correspondingly, the variation of the current in the observed unit amounts at 0.5~mA. 

It is worth considering that in practical operation the set point of each unit is usually varied by amounts much smaller than the total allowed range, so that the absolute cross-talking effect produces disturbances much below the stability and the resolution range.

\section{Test in operative conditions}
A set of eight units was implemented and applied to cancel environmental magnetic field and field gradients in an experiment of high sensitivity field measurement, based on an optical atomic magnetometer \cite{bevilacqua_apb_16}. More precisely, the optical magnetometer operates in a magnetically unshielded room where two components of the environmental magnetic field are fully compensated and a third one is attenuated down to a few $\mu$T, thus it does not provide absolute readings of the Earth magnetic field. The best performance is guaranteed in highly homogeneous magnetic environment, thus five other units supply electro-magnetic quadrupoles for magnetic field inhomogeneity cancellation \cite{belfi_rsi_10}. The magnetometer measures the variations of the environmental magnetic field together with those produced by the imperfections of the magnetic compensation system. Nevertheless, such an experiment can verify the effectiveness of the compensation system, because the magnetic field variations produced by far located sources can be monitored by means of independent data registered and published by several geomagnetic stations over the territory of Italy \cite{geoItaly}. The comparison makes possible to extract information about local magnetic field variations, which include the ones generated by the compensation system. In fact, a large amount the observed environmental field fluctuations is due to far located sources, and they can be effectively cancelled by subtracting signals recorded by other stations as far as hundreds of km. It is worth mentioning that the comparison of magnetic signals recorded by unshielded magnetometers in different locations around the world has implications also in fundamental research \cite{kowalska_arx_16}.

\begin{figure} [htbp]\centering
\includegraphics [width=0.9\columnwidth] {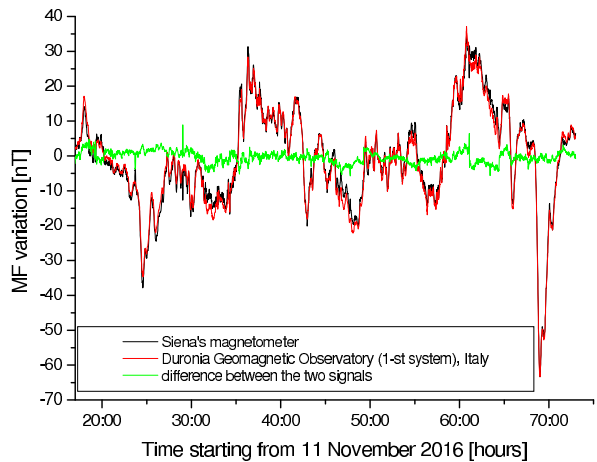}
\caption{Comparison of magnetic field variation as recorded by Duronia geomagnetic station (kind permission of INGV - Geomagnetism Program), and in Siena along two days in November 2016. 
\label{fig:variations}}%
\end{figure}

Fig.\ref{fig:variations} shows the environmental magnetic field variations registered by  our atomic magnetometer during an interval of 60 hours in November 2016. 
 
The scalar nature of our atomic magnetometer makes it sensitive to fluctuations in a direction of the magnetic field, while the geomagnetic observatories in Italy provide vectorial data - the three orthogonal components. In general, the residual magnetic field where the optical magnetometer operates is oriented at a given angle with respect to the main geomagnetic directions. Thus, the evaluation of the fluctuation difference requires combining linearly the three components given by the geomagnetic station in a way to reproduce the variation measured by the optical magnetometer. The red curve in Fig.\ref{fig:variations} shows the linear combination of the three components given by Duronia's geomagnetic station which fits our data (black line). The green curve is the difference between the two magnetometers' readings. Assuming that the readings of the Duronia's geomagnetic station are ideal, the non-zero difference is mainly attributed to the presence of both closely-located field sources and to drift in our eight units magnetic field compensation system. As our magnetometer works in unshielded room, we can not distinguish between these two noises. Assuming the contribution of closely-located sources negligible we can conclude that the residual signal with a standard deviation of 1.8~nT (50~$\mu$A in terms of compensation current) is mainly due to the magnetic field compensation uncertainty. Among the eight units the most important is the one which determines the direction of the magnetometer working (bias) field as along this direction we have the maximum sensitivity. Thus, 50~$\mu$A of uncertainty over 0.5~A of compensation current  corresponds to an upper limit of relative uncertainty of $10^{-4}$ which is consistent with the DAC measured long-term uncertainty.
The mentioned 1.8~nT residual standard deviation can be compared with both the absolute strength of the Earth magnetic field (the Duronia data provide an average value of 46~$\mu$T) and with its typical fluctuation, which in the case of the traces shown in Fig.~\ref{fig:variations} amounts at 13~nT as a root-mean-square value.

\begin{acknowledgments}
This work was partially supported by the national project FIRB RBAP11ZJFA -005, financed by the Italian Ministry for Education, University and Research. 
The authors thank the Italian National Institute for Geophysics and Volcanology (INGV - Geomagnetism Program) for providing their measurements online.

\end{acknowledgments}


\bibliographystyle{apsrev}
\bibliography{v2i}

\begin{thebibliography}{14}
\expandafter\ifx\csname natexlab\endcsname\relax\def\natexlab#1{#1}\fi
\expandafter\ifx\csname bibnamefont\endcsname\relax
  \def\bibnamefont#1{#1}\fi
\expandafter\ifx\csname bibfnamefont\endcsname\relax
  \def\bibfnamefont#1{#1}\fi
\expandafter\ifx\csname citenamefont\endcsname\relax
  \def\citenamefont#1{#1}\fi
\expandafter\ifx\csname url\endcsname\relax
  \def\url#1{\texttt{#1}}\fi
\expandafter\ifx\csname urlprefix\endcsname\relax\def\urlprefix{URL }\fi
\providecommand{\bibinfo}[2]{#2}
\providecommand{\eprint}[2][]{\url{#2}}

\bibitem[{\citenamefont{Skyba}(1991)}]{skyba_rsi_91}
\bibinfo{author}{\bibfnamefont{P.}~\bibnamefont{Skyba}},
  \bibinfo{journal}{Review of Scientific Instruments}
  \textbf{\bibinfo{volume}{62}}, \bibinfo{pages}{2666} (\bibinfo{year}{1991}),
  \urlprefix\url{http://scitation.aip.org/content/aip/journal/rsi/62/11/10.1063/1.1142197}.

\bibitem[{\citenamefont{Burin and Pfleiderer}(1996)}]{burin_rsi_96}
\bibinfo{author}{\bibfnamefont{P.~P.} \bibnamefont{Burin}} \bibnamefont{and}
  \bibinfo{author}{\bibfnamefont{C.}~\bibnamefont{Pfleiderer}},
  \bibinfo{journal}{Review of Scientific Instruments}
  \textbf{\bibinfo{volume}{67}}, \bibinfo{pages}{4023} (\bibinfo{year}{1996}),
  \urlprefix\url{http://scitation.aip.org/content/aip/journal/rsi/67/11/10.1063/1.1147269}.

\bibitem[{\citenamefont{Linzen et~al.}(2004)\citenamefont{Linzen, Robertson,
  Hime, Plourde, Reichardt, and Clarke}}]{linzen_rsi_04}
\bibinfo{author}{\bibfnamefont{S.}~\bibnamefont{Linzen}},
  \bibinfo{author}{\bibfnamefont{T.~L.} \bibnamefont{Robertson}},
  \bibinfo{author}{\bibfnamefont{T.}~\bibnamefont{Hime}},
  \bibinfo{author}{\bibfnamefont{B.~L.~T.} \bibnamefont{Plourde}},
  \bibinfo{author}{\bibfnamefont{P.~A.} \bibnamefont{Reichardt}},
  \bibnamefont{and} \bibinfo{author}{\bibfnamefont{J.}~\bibnamefont{Clarke}},
  \bibinfo{journal}{Review of Scientific Instruments}
  \textbf{\bibinfo{volume}{75}}, \bibinfo{pages}{2541} (\bibinfo{year}{2004}),
  \urlprefix\url{http://scitation.aip.org/content/aip/journal/rsi/75/8/10.1063/1.1771499}.

\bibitem[{\citenamefont{{Belfi} et~al.}(2010)\citenamefont{{Belfi},
  {Bevilacqua}, {Biancalana}, {Cecchi}, {Dancheva}, and {Moi}}}]{belfi_rsi_10}
\bibinfo{author}{\bibfnamefont{J.}~\bibnamefont{{Belfi}}},
  \bibinfo{author}{\bibfnamefont{G.}~\bibnamefont{{Bevilacqua}}},
  \bibinfo{author}{\bibfnamefont{V.}~\bibnamefont{{Biancalana}}},
  \bibinfo{author}{\bibfnamefont{R.}~\bibnamefont{{Cecchi}}},
  \bibinfo{author}{\bibfnamefont{Y.}~\bibnamefont{{Dancheva}}},
  \bibnamefont{and} \bibinfo{author}{\bibfnamefont{L.}~\bibnamefont{{Moi}}},
  \bibinfo{journal}{Review of Scientific Instruments}
  \textbf{\bibinfo{volume}{81}}, \bibinfo{pages}{065103}
  (\bibinfo{year}{2010}).

\bibitem[{\citenamefont{Talukdar et~al.}(2011)\citenamefont{Talukdar,
  Chakraborty, Bose, and Bardhan}}]{talukdar_rsi_11}
\bibinfo{author}{\bibfnamefont{D.}~\bibnamefont{Talukdar}},
  \bibinfo{author}{\bibfnamefont{R.~K.} \bibnamefont{Chakraborty}},
  \bibinfo{author}{\bibfnamefont{S.}~\bibnamefont{Bose}}, \bibnamefont{and}
  \bibinfo{author}{\bibfnamefont{K.~K.} \bibnamefont{Bardhan}},
  \bibinfo{journal}{Review of Scientific Instruments}
  \textbf{\bibinfo{volume}{82}}, \bibinfo{eid}{013906} (\bibinfo{year}{2011}),
  \urlprefix\url{http://scitation.aip.org/content/aip/journal/rsi/82/1/10.1063/1.3509385}.

\bibitem[{\citenamefont{Scandurra et~al.}(2014)\citenamefont{Scandurra,
  Cannat\`a , Giusi, and Ciofi}}]{scandurra_rsi_14}
\bibinfo{author}{\bibfnamefont{G.}~\bibnamefont{Scandurra}},
  \bibinfo{author}{\bibfnamefont{G.}~\bibnamefont{Cannat\`a }},
  \bibinfo{author}{\bibfnamefont{G.}~\bibnamefont{Giusi}}, \bibnamefont{and}
  \bibinfo{author}{\bibfnamefont{C.}~\bibnamefont{Ciofi}},
  \bibinfo{journal}{Review of Scientific Instruments}
  \textbf{\bibinfo{volume}{85}}, \bibinfo{eid}{125109} (\bibinfo{year}{2014}),
  \urlprefix\url{http://scitation.aip.org/content/aip/journal/rsi/85/12/10.1063/1.4903355}.

\bibitem[{\citenamefont{Bevilacqua
  et~al.}(2016{\natexlab{a}})\citenamefont{Bevilacqua, Biancalana, and
  Dancheva}}]{bevilacqua_pra_16}
\bibinfo{author}{\bibfnamefont{G.}~\bibnamefont{Bevilacqua}},
  \bibinfo{author}{\bibfnamefont{V.}~\bibnamefont{Biancalana}},
  \bibnamefont{and} \bibinfo{author}{\bibfnamefont{Y.}~\bibnamefont{Dancheva}},
  \bibinfo{journal}{Phys. Rev. A} \textbf{\bibinfo{volume}{94}},
  \bibinfo{pages}{012501} (\bibinfo{year}{2016}{\natexlab{a}}),
  \urlprefix\url{http://link.aps.org/doi/10.1103/PhysRevA.94.012501}.

\bibitem[{\citenamefont{Bevilacqua
  et~al.}(2016{\natexlab{b}})\citenamefont{Bevilacqua, Biancalana, Chessa, and
  Dancheva}}]{bevilacqua_apb_16}
\bibinfo{author}{\bibfnamefont{G.}~\bibnamefont{Bevilacqua}},
  \bibinfo{author}{\bibfnamefont{V.}~\bibnamefont{Biancalana}},
  \bibinfo{author}{\bibfnamefont{P.}~\bibnamefont{Chessa}}, \bibnamefont{and}
  \bibinfo{author}{\bibfnamefont{Y.}~\bibnamefont{Dancheva}},
  \bibinfo{journal}{Applied Physics B} \textbf{\bibinfo{volume}{122}},
  \bibinfo{pages}{1} (\bibinfo{year}{2016}{\natexlab{b}}), ISSN
  \bibinfo{issn}{1432-0649},
  \urlprefix\url{http://dx.doi.org/10.1007/s00340-016-6375-2}.

\bibitem[{\citenamefont{{Bevilacqua, Giuseppe}
  et~al.}(2016)\citenamefont{{Bevilacqua, Giuseppe}, {Biancalana, Valerio},
  {Ben Amar Baranga, Andrei}, {Dancheva, Yordanka}, and {Rossi,
  Claudio}}}]{bevilacqua_jmr_16}
\bibinfo{author}{\bibnamefont{{Bevilacqua, Giuseppe}}},
  \bibinfo{author}{\bibnamefont{{Biancalana, Valerio}}},
  \bibinfo{author}{\bibnamefont{{Ben Amar Baranga, Andrei}}},
  \bibinfo{author}{\bibnamefont{{Dancheva, Yordanka}}}, \bibnamefont{and}
  \bibinfo{author}{\bibnamefont{{Rossi, Claudio}}}, \bibinfo{journal}{Journal
  of Magnetic Resonance} \textbf{\bibinfo{volume}{263}}, \bibinfo{pages}{65}
  (\bibinfo{year}{2016}), \urlprefix\url{http://arxiv.org/pdf/1510.06250.pdf}.

\bibitem[{\citenamefont{Keeping}(2012)}]{keeping_digikey_12}
\bibinfo{author}{\bibfnamefont{S.}~\bibnamefont{Keeping}},
  \bibinfo{journal}{DigiKey}  (\bibinfo{year}{2012}),
  \urlprefix\url{http://www.digikey.com/en/articles/techzone/2012/may/hybrid-power-supplies-deliver-noise-free-voltages-for-sensitive-circuitry}.

\bibitem[{\citenamefont{Teel}(2015)}]{teel_ti_05}
\bibinfo{author}{\bibfnamefont{J.~C.} \bibnamefont{Teel}},
  \bibinfo{journal}{Texas Instrum Appl. Notes}  (\bibinfo{year}{2015}),
  \urlprefix\url{http://www.ti.com/lit/an/slyt201/slyt201.pdf}.

\bibitem[{\citenamefont{Keeping}(2015)}]{keeping_digikey_15}
\bibinfo{author}{\bibfnamefont{S.}~\bibnamefont{Keeping}},
  \bibinfo{journal}{DigiKey}  (\bibinfo{year}{2015}),
  \urlprefix\url{http://www.digikey.com/en/articles/techzone/2015/jul/understanding-linear-regulator-noise-in-hybrid-power-supplies}.

\bibitem[{geo()}]{geoItaly}
\urlprefix\url{http://geomag.rm.ingv.it}.

\bibitem[{\citenamefont{{Kowalska-Leszczynska}
  et~al.}(2016)\citenamefont{{Kowalska-Leszczynska}, {Bizouard}, {Bulik},
  {Christensen}, {Coughlin}, {Go{\l}kowski}, {Kubisz}, {Kulak}, {Mlynarczyk},
  {Robinet} et~al.}}]{kowalska_arx_16}
\bibinfo{author}{\bibfnamefont{I.}~\bibnamefont{{Kowalska-Leszczynska}}},
  \bibinfo{author}{\bibfnamefont{M.-A.} \bibnamefont{{Bizouard}}},
  \bibinfo{author}{\bibfnamefont{T.}~\bibnamefont{{Bulik}}},
  \bibinfo{author}{\bibfnamefont{N.}~\bibnamefont{{Christensen}}},
  \bibinfo{author}{\bibfnamefont{M.}~\bibnamefont{{Coughlin}}},
  \bibinfo{author}{\bibfnamefont{M.}~\bibnamefont{{Go{\l}kowski}}},
  \bibinfo{author}{\bibfnamefont{J.}~\bibnamefont{{Kubisz}}},
  \bibinfo{author}{\bibfnamefont{A.}~\bibnamefont{{Kulak}}},
  \bibinfo{author}{\bibfnamefont{J.}~\bibnamefont{{Mlynarczyk}}},
  \bibinfo{author}{\bibfnamefont{F.}~\bibnamefont{{Robinet}}},
  \bibnamefont{et~al.}, \bibinfo{journal}{ArXiv e-prints}
  (\bibinfo{year}{2016}), \eprint{1612.01102}.

\end{thebibliography}

\end{document}